\documentclass[aps,reprint,superscriptaddress,nofootinbib]{revtex4-2}

\usepackage{amsmath,amssymb,graphicx,bm}
\usepackage{hyperref}
\usepackage{float}
\usepackage{xcolor,verbatim}
\usepackage{bm}
\usepackage{amsfonts}
\usepackage{latexsym}
\usepackage{physics}
\usepackage{comment,mathtools}

\newcommand{\be}{\begin{equation}}
\newcommand{\ee}{\end{equation}}
\def\sfrac#1#2{{\textstyle{#1\over #2}}}
\newcommand{\bea}{\begin{eqnarray}}
\newcommand{\eea}{\end{eqnarray}}
\newcommand{\nn}{\nonumber}

\begin{document}

\title{Dark Photons from Perturbative Decay of a Misaligned Higgs Field}
\author{James M.\ Cline}
\author{Gonzalo Herrera}
\author{Jean-Samuel Roux}

\affiliation{McGill University Department of Physics \& Trottier Space Institute, 3600 University St., Montr\'eal, QC H3A 2T8 Canada}

\begin{abstract}
We reconsider the production of dark photons  $A'$ as dark matter, from the perturbative decay of a dark Higgs field $h$, that is stochastically misaligned from the minimum of its potential during inflation.  This is a simple and predictive framework for generating the $A'$ relic abundance.
It is constrained by structure formation, since the $A'$ are initially boosted, and inflationary isocurvature fluctuations, which require small quartic couplings $\lambda h^4$.  We identify $A'$ masses
between 100\,eV and 1\,GeV and gauge couplings $g\sim 10^{-15}-10^{-10}$ that are consistent in this scenario, and which become more tightly constrained if a generic level of kinetic mixing is present. The favored parameter region could be tested through future CMB or Lyman-$\alpha$ observations, and, in the presence of kinetic mixing, by direct detection experiments or diffuse soft gamma-ray searches.
\end{abstract}

\maketitle

\section{Introduction}
Dark photons ($A'$s) have been highly scrutinized as a dark matter (DM)
candidate, both theoretically and in experimental searches.  Similarly
to axion-like particles, their masses can be very small, well below
that of thermal relics, making them a natural candidate for light
DM.  

Unlike axions,  the misalignment mechanism does not naturally
explain the relic density of dark photons, owing to their conformally
invariant kinetic term.  One must significantly complicate the model
with nonminimal couplings to gravity to achieve the relic density from
displacement of the $A'$ in its potential \cite{Arias:2012az}.  Another 
possibility  is $A'$ production from inflationary perturbations, or
gravitational production \cite{Graham:2015rva}. This mechanism can be effective if the $A'$ mass
is larger than $\sim 10^{-5}$ eV and generated from the Stueckelberg
mechanism, treating the mass as a fundamental parameter in the Lagrangian of the
theory. However, if the $A'$ mass is generated by the Higgs
mechanism and is therefore field-dependent, gravitational production is typically suppressed by the increased
$m_{A'}$ mass during inflation, making
its contribution negligible for our purposes \cite{Sato:2022jya}.

Alternatively, dark photons could inherit their
abundance from a misalignment of the dark Higgs field $h$ that gives
mass to the $A'$, and perturbatively decays into it
via $h\to A'A'$.  Such a misalignment could naturally be
generated during inflation, by the stochastic build-up of $h$
fluctuations.
The production of $A'$s from dark Higgs parametric resonance was studied in Ref.\ \cite{Dror:2018pdh}, treating the degree of Higgs misalignment as a free parameter.
Ref.\ \cite{Cline:2024wja} (see also \cite{Redi:2022zkt, Khan:2026bmf, Khan:2026php}) pointed out that in fact there is a distribution of probable values of the Higgs misalignment that is generated during inflation, which is correlated with the scale of inflation,
and that the Higgs decays can be perturbative, 
making this scenario more predictive. 

In the present paper, we explore this appealing production mechanism more quantitatively, with a careful study of the $h$ evolution after its inflationary stochastic misalignment, that depends upon whether the $h^4$ or $h^2$ terms in its potential are dominating.  Taking account of constraints from structure formation and isocurvature perturbations, we identify an interesting viable window for $A'$
masses $100\,{\rm eV}\lesssim m_{A'}\lesssim 1\,$GeV and gauge couplings
$10^{-15}\lesssim g\lesssim 10^{-10}$ in the absence of kinetic mixing.  If the kinetic mixing takes its generic
loop-induced value $\epsilon \sim e g/(16\pi^2)$, the allowed region becomes more restricted, $10\,{\rm keV}\lesssim m_{A'} \lesssim 1\,$MeV, for $10^{-17}\lesssim \epsilon \lesssim 10^{-14}$.

In the following, we describe our framework in section \ref{sec:setup}, and in section \ref{sec:misalignment} we describe the stochastic misalignment formalism and isocurvature perturbations produced during inflation. In section \ref{sec:fieldevol}, we present the field evolution of the dark Higgs after inflation and its decay into dark photons. In section \ref{sec:dm}, we constrain the dark photon's relic density and its cooling prior to structure formation. In section \ref{sec:inf}, we discuss other constraints on our model, namely the weak gravity conjecture and alternative production mechanisms during inflation. In section \ref{sec:epsilon}, we recast our bounds in terms of the kinetic mixing parameter, assuming the dark photon mixes with the visible photon at the one-loop level.  Conclusions are given in section \ref{sec:conc}.

\section{Theoretical framework}
\label{sec:setup}
We initially consider the dark photon plus Higgs Lagrangian in the absence of kinetic mixing,
\be \label{eq:lagrangian}
\mathcal{L}= -\frac14 F_{\mu\nu}'F^{\prime \mu\nu}+ {\big|}D_\mu H{\big|}^2 - V(|H|)
\ee 
with scalar potential
\be \label{eq:pot}
V(|H|) = \lambda \left(|H|^2-\frac{1}{2} v^2\right)^2
\ee 
and covariant derivative $D_\mu = \partial_\mu - ig A'_\mu$.  The complex Higgs is decomposed as $H = \sfrac{1}{\sqrt{2}} h e^{i\theta}$, where the Goldstone boson $\theta$ is eaten by the $A'$ gauge boson.

During inflation, the dark Higgs will be displaced away from its vacuum expectation value (VEV) $v$ by stochastic fluctuations \cite{Starobinsky:1994bd,Graham:2018jyp}
to some characteristic value $h_0$, to be treated in greater detail in section \ref{sec:misalignment}.  Here we simply take $h_0$ to be an initial value, from which $h$ eventually evolves toward $v$.

During intermediate times, the masses of $h$ and $A'$ are field-dependent,
\bea
m_h^2(h) &=& \lambda h^2 \left(3 - \frac{v^2}{h^2}\right) \label{eq:mh}\\
m_{A'}(h)&=& g \abs{h}
\eea
Once $h$ settles to its VEV, the two masses are related by
\be 
\frac{m_{A'}}{m_{h}}= \frac{g}{\sqrt{2\lambda}}.
\label{massratio}
\ee
Therefore the decay of $h$ into two dark photons occurs only if $g < \sqrt{\lambda/2}$.
Later we will consider the situation where kinetic mixing between $A_\mu'$ and the SM photon arises at one loop, $\epsilon\sim ge/(16\pi^2)$. In this case, experimental constraints on $\epsilon$ will lead to $g\ll\sqrt{\lambda}$, hence $m_{A'}\ll m_h$.

\section{Stochastic misalignment and isocurvature perturbations}
\label{sec:misalignment}
Here we will review the build-up of Higgs misalignment during inflation, and a stringent bound on its self-coupling arising from the associated isocurvature fluctuations.
Stochastic misalignment is characterized by a
correlation between the Hubble rate during inflation, $H_I$ and the typical field displacement $h_0$ by the end of inflation. 

If inflation lasts for sufficiently many $e$-foldings, the classical slow-roll of the background complex field $H$ is countered by the quantum fluctuations of modes that exit the horizon. This results in a random walk where the field value inside a given Hubble volume follows a probability distribution function  \cite{Starobinsky:1994bd}
\be 
f(H) \propto \exp\!\left(-\frac{8\pi^2 V(H)}{3 H_I^4}\right).
\ee 
For the potential of Eq.~(\ref{eq:pot}), this becomes
\be \label{eq:pdfcomplex}
f(h,\theta) \propto \exp\!\left(-\frac{2 \pi^2 \lambda}{3 H_I^4} \left(h^2-v^2\right)^2 \right)\,,
\ee 
assuming that the dark Higgs is light compared to $H_I$.

Eq.\ (\ref{eq:pdfcomplex}) is a two-dimensional random distribution with a characteristic width $\Delta_h\sim H_I/\lambda^{1/4}$, that has two interesting limits. For $\Delta_h\ll v$, the dark Higgs amplitude has a nearly Gaussian distribution centered around $v$. In this case, it is convenient to expand the scalar field as $h=v+\phi$, which has an approximately quadratic potential $V(\phi)\approx \lambda v^2\phi^2$ from Eq.~(\ref{eq:pot}). In the opposite limit $\Delta_h \gg v$, the VEV is negligible and $h$'s
distribution function goes as $\sim h^4$ in the exponent.  Henceforth we refer to these cases as
the quadratic or quartic regime, respectively.
The regime in which the field initially oscillates is determined by whether $\lambda v^4/H_I^4 < 1$, corresponding to quartic,
or $\lambda v^4/H_I^4 > 1$, giving quadratic.
Quartic oscillations eventually give rise to quadratic ones as the amplitude  Hubble-damps, but we make the distinction to indicate the regime in which the field starts at the end of inflation.

We derive in Appendix \ref{app:PDF} the following 
mean values and standard deviations for the 
radial $h$ distribution function:
\begin{align}\label{eq:meanh}
\ev{h}&\simeq \left\lbrace \begin{array}{lr}
     v&  \text{(quadratic)}\\
     \dfrac{\Gamma(3/4)}{\sqrt{\pi}}\Delta_h& \text{(quartic)} 
\end{array}
\right.\\
\sigma&\simeq \left\lbrace \begin{array}{lr} \label{eq:stdh}
     \sqrt{\dfrac{\Delta_h^4}{8v^2}}&  \text{(quadratic)}\\
     \left(\dfrac{1}{\sqrt{\pi}} - \dfrac{\Gamma(3/4)^2}{\pi}\right)^{1/2}\Delta_h& \text{(quartic)} 
\end{array}
\right.
\end{align}
where $\Delta^4_h \equiv 3H_I^4/(2\pi^2\lambda)$. 
These determine the likelihood of a given initial amplitude, understood to be the mean value inside of a Hubble volume.
However, fluctuations of $h$ on sub-Hubble scales decay into spatially inhomogeneous $A'$s, which appear as isocurvature fluctuations in the cosmic microwave background (CMB). The {\it Planck} collaboration derives a strict upper bound on the isocurvature fraction parameter, $\beta_{\rm iso}=\mathcal{P}_S/(\mathcal{P}_S+\mathcal{P}_R)<0.038$ \cite{Planck:2018jri}, where $\mathcal{P}_S$ and  $\mathcal{P}_R$ are the isocurvature and adiabatic power spectra, respectively. 

Suppression of the isocurvature fluctuations requires a large initial amplitude. Following Ref.~\cite{Khan:2026bmf} we show in Appendix \ref{app:iso} that the Planck limit can be interpreted as a lower bound on the field misalignment during inflation,
\be \label{eq:isoField}
\begin{array}{lr} %qeff * 1.747455711e4
  \phi_0 \gtrsim 3.5\times 10^{4}\ f_{A'} H_I\quad &\text{(quadratic),}\\
    h_0 \gtrsim 5.2\times 10^{4}\ f_{A'} H_I \quad  &\text{(quartic)},
\end{array}
\ee 
where $f_{A'}=\Omega_{A'}/\Omega_{\rm DM}$ is the fraction of dark matter made up of dark photons.
The initial field value should not exceed the mean values given in Eq.~(\ref{eq:meanh}) by more than a few $\sigma$ from Eq.\ (\ref{eq:stdh}).  Requiring no more than a $2\sigma$ excursion implies
\be \label{eq:isoParam}
\begin{array}{rlr} 
  m_h=\sqrt{2\lambda}v& \lesssim 1.1\times 10^{-5}\ f_{A'}^{-1}\,H_I\quad &\text{(quadratic),}\\ %1.115474339e-5
    \lambda &\lesssim 5.4 \times10^{-20}  f_{A'}^{-4}\,\quad  &\text{(quartic)}. %5.377437925e-20
\end{array}
\ee 
 
 In the quadratic regime, combining the first bound in (\ref{eq:isoParam}) with the consistency condition $\Delta_h\lesssim v$ leads to the upper bound $\lambda\lesssim2.5\times10^{-20} f_{A'}^{-4}$, in agreement with Ref.~\cite{Khan:2026bmf}.\footnote{Ref.~\cite{Redi:2022zkt} performed a similar analysis, but incorrectly quoted the bound $\lambda\lesssim10^{-10}$ for the quartic scenario.} 
 Such a small value for the quartic coupling is consistent with quantum corrections as long as the gauge coupling is also small. The dark photon loop contributes a factor
 \be 
\delta\lambda \simeq \frac{g^4}{16\pi^2}
 \ee 
to the dark Higgs self-interaction. Given that perturbative decays require $g<\sqrt{\lambda/2}\sim10^{-10}$, this gives $\delta \lambda\sim 10^{-42}$ for our parameters of interest, which is negligible.

Refs.~\cite{Redi:2022zkt,Markkanen:2018gcw} showed that the opposite regimes, $m_h\sim H_I$ or $\lambda\gtrsim1$, could also result in an exponentially suppressed isocurvature power spectrum. However, those parameter choices are incompatible with our scenario: misalignment requires $m_h\ll H_I$ in order for the dark Higgs to be frozen during inflation, and for a large value of $\lambda$ the initial amplitude $h_0$ required to explain the dark matter relic density would put $V(h_0)$ above the inflationary scale (and its upper limit implied by the tensor-to-scalar ratio \cite{Planck:2018jri}; see Eq.~(\ref{eq:phi0DM})). Therefore, we will focus on the regions described in Eq.~(\ref{eq:isoParam}). 

The equilibrium distribution function Eq.~(\ref{eq:pdfcomplex}) is only reached after many $e$-foldings, much more than the minimum number $N\sim50-60$ required for inflation. The stochastic distribution relaxes to its equilibrium for $N$ of order \cite{Starobinsky:1994bd, Redi:2022zkt}
\be 
N_{\rm rel}\simeq \left\lbrace \begin{array}{lr}
     \dfrac{3H_I^2}{2 m_{h}^2}  \quad &\text{(quadratic),}\\
     \sqrt{\dfrac{8\pi^2}{9 \lambda}}\quad &\text{(quartic).} 
\end{array}\right.
\ee 
Using the isocurvature bound Eq.~(\ref{eq:isoParam}), we find $N_{\rm rel}\sim \mathcal{O}(10^{10})$ is required to be consistent with stochastic inflation. Such a long period of inflation is also sometimes encountered in axion models and relaxion solutions to the hierarchy problem \cite{Graham:2015cka,Graham:2018jyp}.

Slow-roll inflation cannot last for an arbitrary long period of time. If the inflaton fluctuations dominate over the classical field evolution, inflation will be eternal in some parts of the universe. To prevent this, the total number of $e$-foldings must satisfy  \cite{Dubovsky:2011uy}
\be 
N< \frac{2\pi^2 M_p^2}{3 H_I^2}.
\ee 
Requiring $N_{\rm rel}\sim 10^{10}$ yields the upper bound $H_I\lesssim 6\times 10^{13}$ GeV, which accidentally coincides with upper bound on the inflationary scale set by \textit{Planck} \cite{Planck:2018jri}. We will show in section \ref{sec:dm} that a Hubble scale $H_I\sim 10^{10}$ GeV or lower is sufficient for the dark Higgs misalignment to yield the dark matter relic density in the favored region of parameter space. Therefore, our mechanism is  consistent with the requirements of stochastic misalignment.

 \section{Scalar field evolution}
 \label{sec:fieldevol}
 Having established the likely initial condition $h_0$ of the dark Higgs starting from the end of inflation, we now study its subsequent evolution
 before decaying into the dark photons.  Although oscillations of $h$ could produce $A'$s through parametric resonance, this is inefficient 
 when $g\ll\lambda$, which is the relevant regime for our study.  Therefore perturbative $h\to A'A'$ decays are the dominant source of
 dark photons.  A detailed discussion of alternative production mechanisms is given in 
 Section \ref{sec:gravProd}.

Once $H$ falls below the field-dependent $m_h$ the scalar field will start to oscillate. The evolution of the dark Higgs falls into one of three categories. (i) If $h_0\gg v$, the potential is effectively quartic during the oscillations and when decays occur, so that $V(h)\approx\sfrac{1}{4}\lambda h^4$. (ii)  If $h_0\lesssim v$, the potential is approximately quadratic at all later times, $V(\phi)\approx \lambda v^2 \phi^2$.  (iii) If $h_0\gtrsim v$, the field will be in a hybrid regime where the effective potential is quartic at first, but the field evolves to the quadratic regime and oscillates about $h=v$ before it decays.  
Which of these three regimes applies is determined by $\lambda$, $g$ and $m_{A'}$ see Eq.\ (\ref{eq:regimesmA}) in section \ref{sec:dm}.

The $h$ oscillations begin once $H\sim m_h$. It is convenient to use the following approximation to analyze the field evolution in each regime,
\be
m_h \simeq \left\lbrace\begin{array}{lr}
    \sqrt{3\lambda}\,h &\quad \text{(quartic, hybrid)},\\
    \sqrt{2\lambda}\,v &\text{(quadratic)}.
\end{array}\right.
\ee
One can write the VEV as $v=m_{A'}/g$, where $m_{A'}$ is the late-time mass of the dark photon. The temperature at which oscillations begin can then be estimated as 
\bea \label{eq:Tosc}
T_{\rm osc}&\simeq& \left(\frac{90\, m_h^2 M_p^2}{\pi^2 g_{*}}\right)^{1/4} \\% 
\nn&\simeq&\left\lbrace\begin{array}{lr}
     0.71\, \lambda^{1/4} (h_0 M_p)^{1/2}  &\quad \text{(quartic, hybrid)},\\%0.7114985736
    0.64\, \lambda^{1/4} \left(\dfrac{m_{A'} M_p}{g}\right)^{1/2} &\text{(quadratic)}. %0.6429115367
\end{array}\right.
\eea 
where $M_p=2.435\times10^{18}$ GeV is the reduced Planck mass and we used $g_*=106.75$, since our constraints will eventually imply that
$T_{\rm osc}$ is above the electroweak scale.

Decays of $h$ occur when $H\sim \Gamma$, at the temperature $T_d$. For large field values $h\gg v$, the decay rate is field-dependent \cite{Araki:2020wkq,Djouadi:2005gi},
\be
\Gamma = \frac{m_h^3}{32\pi h^2}\sqrt{1-4x}(1-4x+12x^2)
\ee 
where $x=m_{A'}^2/m_h^2$. % 
In the limit $x\to0$ we find
\be 
\Gamma \simeq \frac{m_h^3}{32\pi h^2} = \frac{\lambda^{3/2} h}{32 \pi} \left(3-\frac{v^2}{h^2}\right)^{3/2} . 
\ee
 If $h$ is initially in the quartic regime, its amplitude redshifts as $h\propto a^{-1}$, and so does $\Gamma$. But during radiation domination the Hubble parameter decreases faster, $H\propto a^{-2}$. Therefore $h$ may decay before its amplitude becomes small enough to oscillate about its VEV. 

If the decays do not occur during the quartic regime, the field begins to oscillate about its VEV once $h \sim v$. The potential becomes effectively quadratic, and the decay rate is constant; this is the hybrid regime. The transition from quartic to quadratic oscillations takes place at a temperature
\be 
T_v = \frac{v}{h_0} T_{\rm osc}.
\ee
If $h_0\sim v$, this gives $T_v\simeq T_{\rm osc}$ and the field behaves quadratically from the outset. Recall that in this case, $h=v+\phi$ such that $V\simeq\sfrac{1}{2}m_h^2\phi^2$, valid as long as  $|\phi|\ll v$.

On the other hand, if $h_0\gtrsim32 \pi v/3\lambda$, the Hubble parameter reaches $H\sim \Gamma$ before $T_v$, \textit{i.e.,} the dark Higgs decays in the quartic regime. The initial conditions therefore determine in which regime the oscillations take place. In terms of the free parameters, the  boundaries are approximately
\begin{equation} \label{eq:cond}
        \left.\begin{array}{ll}
               h_0 \gtrsim \dfrac{32\pi}{3\lambda} \dfrac{m_{A'}}{g}\quad&\Rightarrow {\rm quartic};\\ \\
              |\phi_0|\lesssim \dfrac{m_{A'}}{g}&\Rightarrow {\rm quadratic;}\\
             \text{in between} & \Rightarrow{\rm hybrid}.
              \end{array}\right.
\end{equation}
Eq.~(\ref{eq:cond}) shows that the quadratic regime is typically valid for heavier dark photons or weaker gauge coupling relative to the quartic regime.

The decay temperature also depends on the evolution of $h$,
\be \label{eq:td}
T_d \simeq \left\lbrace\begin{array}{ll}
    \dfrac{3\lambda}{32\pi} T_{\rm osc} &  \text{(quartic)},\\
    \left(\sqrt{\dfrac{2}{3}}\dfrac{\lambda}{16\pi}\dfrac{v}{h_0}\right)^{1/2} T_{\rm osc}\quad &\text{(hybrid)},\\
     \sqrt{\dfrac{\lambda}{16 \pi}}\, T_{\rm osc} & \text{(quadratic)}.
\end{array}\right.
\ee 
Despite appearances, $T_d$ in the hybrid and quadratic scenarios are actually equal in value. The difference lies in the definitions of $T_{\rm osc}$ in the two cases.

\section{Dark photon relic density and coldness constraints}
\label{sec:dm}

The previous determinations allow us to estimate the initial field amplitude required for dark photons to constitute a fraction $f_{A'}$ of dark matter. At matter-radiation equality ($T_{\rm eq}\approx 0.80$ eV), $\rho_{\rm m}=\rho_{\rm r,eq}$ where $\rho_{\rm m}=\rho_c+\rho_b=\rho_c (1+\Omega_b/\Omega_c)$ is the total matter density, including cold dark matter and baryons. Imposing $\rho_{A'}=f_{A'}\rho_c$, the $A'$ abundance follows from the abundance of $h$ when the field starts to oscillate,
\be\label{eq:habundance}
Y_h = \frac{\rho_h}{m_h s_{\rm osc}} = \left\lbrace \begin{array}{ll}
     \dfrac{1}{4}\sqrt{\dfrac{\lambda}{3}}\dfrac{h^3_0}{ s_{\rm osc}} \quad& \text{(quartic, hybrid),}  \\
    \dfrac{\sqrt{2\lambda}}{2}\dfrac{v\phi_0^2}{ s_{\rm osc}} \quad& \text{(quadratic)}, 
\end{array} \right.
\ee
(recall $h=v+\phi$ in the quadratic regime) and $s_{\rm osc}=(2\pi^2/45)g_{s*} T_{\rm{osc}}^3$.\footnote{Ref. \cite{Cline:2024wja} incorrectly evaluated $Y_h$ at the reheating temperature rather than at $T_{\rm osc}$ and therefore underestimated its comoving abundance}.  $Y_h$ is adiabatically conserved prior to the decays, no matter the regime in which the field is oscillating.

\begin{figure*}[!t]
\centerline{
\includegraphics[scale=0.8]{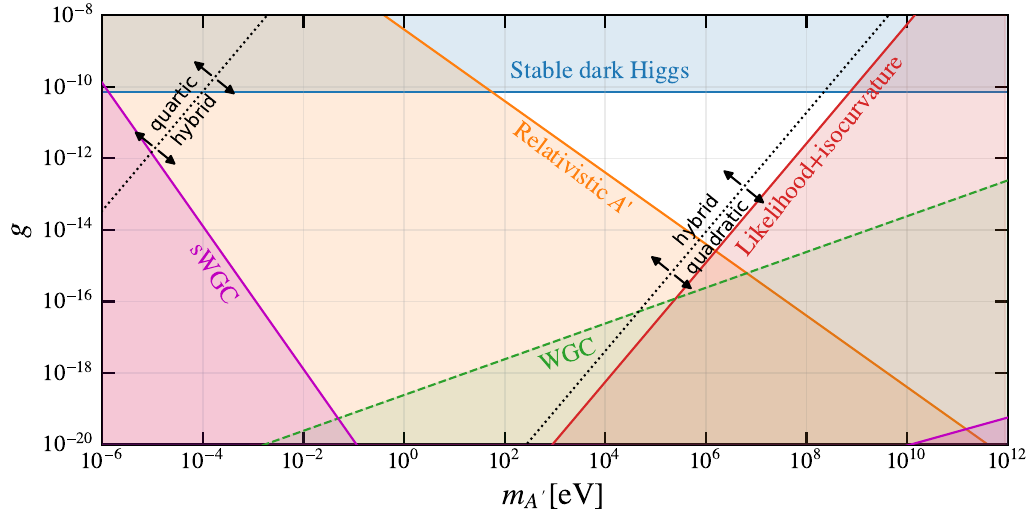}
}
\vspace{-0.2cm}
\caption{Constraints on the dark photons mass versus the hidden sector gauge coupling, assuming $\lambda =10^{-20}$ and $f_{A'}=\Omega_{A'}/\Omega_{\rm DM}=1$. The black dotted lines demarcate the different regimes of oscillation.  Blue: $h\to A'A'$ decays kinematically forbidden. Orange:  dark photons still relativistic during structure formation. Red: initial $h$ misalignment is outside $\pm2\sigma$ of its expected value, given the upper bound on the inflationary Hubble scale set by isocurvature perturbations. Green, dashed: the dark Higgs does not satisfy the weak gravity conjecture, and the addition of a lighter dark-charged species is necessary. Purple:  inflationary Hubble scale is greater than the cutoff imposed by the sublattice weak gravity conjecture. 
 \label{fig:mg}}
\end{figure*}

Assuming $h$ decays only into dark photons, their final abundance is $Y_{A'}= n_{A'}/s=2Y_h$. The $A'$ relic density is then
\be
\rho_{A'}=m_{A'}n_{A',{\rm eq}} = \frac{f_{A'}}{1+\Omega_b/\Omega_c}\, \rho_{\rm r,eq}\,. %
\ee
Using $g_{*,\rm{osc}}=106.75$, $g_{*,\rm{eq}}=3.363$, $g_{s*,\rm{eq}}=3.909$ \cite{Husdal:2016haj}, $\Omega_b/\Omega_c=0.1864$
\cite{Planck:2018vyg} and Eq.~(\ref{eq:Tosc}), we find that
\be \label{eq:phi0DM}
h_0 = 9.8\times10^{15}\ {\rm{GeV}}\ f_{A'}^{2/3}\left(\frac{\lambda}{10^{-20}}\right)^{1/6} \left(\frac{1\ {\rm eV}}{m_{A'}}\right)^{2/3} %9.771286079e15
\ee 
for the quartic and hybrid regimes, and 
\bea  \label{eq:phi0DMquad}
\phi_0&=&1.2\times10^{12}\ {\rm GeV}\ f_{A'}^{1/2}\left(\frac{\lambda}{10^{-20}}\right)^{1/ 8} \left(\frac{{\rm 1\ MeV}}{m_{A'}}\right)^{1/ 4} \nn\\ &&\times\left(\frac{10^{-17}}{g}\right)^{1\over 4} %1.206081800e12
\eea
for the quadratic oscillations. The condition $f_{A'}\leq1$ yields an upper bound on the initial field amplitude. Combining these expressions with Eq.~(\ref{eq:cond}) determines which regime applies, through
\begin{equation} \label{eq:regimesmA}
        \left.\begin{array}{ll}
               \frac{m_{A'}}{f_{A'}^{2/5}} \lesssim7.6\times 10^{-6}\ {\rm eV}\left(\dfrac{\lambda}{10^{-20}}\right)^{\!\!7\over 10} \!\!\left(\dfrac{g}{10^{-12}} \right)^{3\over 5}&\Rightarrow {\rm quartic};\\ \\%7.565957185e-6 eV
              \frac{m_{A'}}{f_{A'}^{2/5}} \gtrsim 17\ {\rm keV}\left(\dfrac{\lambda}{10^{-20}}\right)^{1\over 10} \left(\dfrac{g}{10^{-17}} \right)^{3\over 5}&\!\!\!\!\!\!\Rightarrow {\rm quadratic;}\\ %16.76013122 keV
              \\
              \frac{m_{A'}}{f_{A'}^{2/5}} {\rm \ in\ between} & \Rightarrow{\rm hybrid}\,,
              \end{array}\right.
\end{equation}
These boundaries are shown as dotted lines in Fig.\ \ref{fig:mg} for $\lambda=10^{-20}$ and $f_{A'}=1$. 

\subsection{Cooling of dark photons}

The dark photons produced from $h$ decays are initially relativistic, $p\simeq m_{h,d}/2\gg m_{A'}$, where $m_{h,d}$ is the effective dark Higgs mass at the time of the decays. This could allow them to stream out of overdense regions, suppressing density fluctuations at small scales; hence they must cool sufficiently by the time of structure formation. This makes light $A'$s a particular case of warm dark matter (WDM). For a thermal relic, the matter power spectrum and in particular the Lyman-$\alpha$ forest constrain WDM mass to $m_{\rm WDM}\gtrsim (1.9-5.3)$ keV, with the exact bound depending on the model and assumptions about the thermal history of the intergalactic medium. \cite{Palanque-Delabrouille:2019iyz,Garzilli:2019qki,Villasenor:2022aiy} 

However, $A'$s are non-thermal in our model, so the bounds on $m_{\rm WMD}$ cannot be directly applied to $m_{A'}$. Instead, one should compute the matter spectrum by solving the Boltzmann equation for the phase space distribution $f_{A'}(p,t)$, then compare the result with the thermal relic scenario to find the corresponding value of $m_{\rm WMD}$. This has been done in Ref.~\cite{Ballesteros:2020adh} for various non-thermal production mechanisms. For DM produced from the decay of scalar condensate, we can use the following mapping between the $A'$ mass and its equivalent for a thermal relic WDM,
\be
m_{\rm WDM}^{\rm eq} = 1.1\ {\rm keV}\ \left(\frac{m_{A'}}{1\ {\rm keV}}\right)^{3/4} \left( \frac{T_D}{m_{h,d}}\right)^{3/4}. 
\ee 

Using the conservative bound $m_{\rm WDM}^{\rm eq}\gtrsim 1.9$ keV \cite{Garzilli:2019qki} leads to the following constraint\footnote{ Eq.~(\ref{eq:coldnessBound}) can also be obtained by following Ref.~\cite{Dror:2018pdh} and requiring that $A'$s become non-relativistic at a temperature $T_{\rm nr}\gtrsim4$ keV.}
\be \label{eq:coldnessBound}
m_{A'} \gtrsim \left\lbrace \begin{array}{l}
    0.05\ {\rm eV}\ \left(\dfrac{\lambda}{10^{-20}}\right)^{1/4} \left( \dfrac{h_0}{M_p}\right)^{1/2} \\ \quad \text{(quartic)}, \\%0.01217185018 eV
     4.2\ {\rm MeV}\ \left(\dfrac{10^{-15}}{g}\right) \left(\dfrac{10^{-20}}{\lambda}\right)^{1/2}\\ \quad \text{(hybrid, quadratic)}\,.
\end{array}\right.
\ee 
where we used Eq.~(\ref{eq:td}) and assumed $T_d/h_d\simeq T_{\rm osc}/h_0$ in the quartic scenario.
For quartic oscillations, one can combine this bound with Eq.~(\ref{eq:phi0DM}), assuming the dark photons make up all the dark matter, giving
\be \label{eq:quarticDMCold}
m_{A'} \gtrsim 13\ {\rm meV}\ \left(\frac{\lambda}{10^{-20}}\right)^{\frac14} \text{(quartic + DM relic density)}.%13.36 meV
\ee 
However, this part of parameter space is only consistent with quartic oscillations identified in Eq.~(\ref{eq:regimesmA}) if $g\gtrsim 10^{-7}$. Such values are incompatible with the requirement $m_{A'}\leq m_h/2$ to allow perturbative decays and the isocurvature constraint $\lambda\lesssim\mathcal{O}(10^{-20})$.  We therefore conclude that the fully quartic regime, in which the dark Higgs starts to decay at some amplitude $h\gg v$, cannot consistently produce sufficiently cold $A'$ dark matter.

The hybrid and quadratic regimes remain interesting possibilities. Eq.~(\ref{eq:coldnessBound}) allows dark photon dark matter as light as $m_{A'}\sim 100$ eV. Previously, Ref.~\cite{Dror:2018pdh} concluded that the perturbative decay of a misaligned dark Higgs cannot produce cold dark photon dark matter below the keV scale, but they only considered the case of quadratic oscillations.

Having determined the required dark Higgs amplitude that yields the dark matter relic density, we can combine Eqs.~(\ref{eq:phi0DM}) and (\ref{eq:phi0DMquad}) with our results from section \ref{sec:misalignment} to bound the inflationary Hubble scale. The isocurvature constraint Eq.~(\ref{eq:isoField}) imposes an upper limit on $H_I$,
\be \label{eq:HIupperBound}
H_I \lesssim \left\lbrace\begin{array}{l}
     1.9\times10^{11}\ \text{GeV}\ f_{A'}^{-1/3}\left(\dfrac{\lambda}{10^{-20}}\right)^{1/6}\left(\dfrac{1\ {\rm eV}}{m_{A'}}\right)^{2/3}\\ %1.863907245e11 GeV
     \qquad \text{(quartic, hybrid);}  \\
     3.5\times 10^{7}\ \text{GeV}\ f_{A'}^{-1/2}\left(\dfrac{\lambda}{10^{-20}}\right)^{1/8} \left(\dfrac{{\rm 1\ MeV}}{m_{A'}}\right)^{1/ 4} \\\qquad\times\left(\dfrac{10^{-17}}{g}\right)^{1/4} %3.450965287e7 GeV
     \qquad \text{(quadratic).}
\end{array}\right.
\ee

On the other hand, stochastic misalignment bounds  $H_I$ from below, if dark photons make up a significant fraction of dark matter. The lower $H_I$, the less likely the amplitudes of Eqs.~(\ref{eq:phi0DM}) and (\ref{eq:phi0DMquad}) become. Using Eqs.~(\ref{eq:meanh}) and (\ref{eq:stdh}) and requiring that the initial misalignment lies within $2\sigma$ of the expected field value yields the likelihood constraint,
\be \label{eq:HIlowerBound}
H_I \gtrsim \left\lbrace\begin{array}{l}
     1.2\times10^{11}\ \text{GeV}\ f_{A'}^{2/3}\left(\dfrac{\lambda}{10^{-20}}\right)^{5/12}\left(\dfrac{1\ {\rm eV}}{m_{A'}}\right)^{2/3}\\ %1.223997258e11 GeV
     \qquad \text{(quartic, hybrid);}  \\
     2.1\times 10^{8}\ \text{GeV}\ f_{A'}^{1/4}\left(\dfrac{\lambda}{10^{-20}}\right)^{5/{16}} \left(\dfrac{m_{A'}}{\rm 1\ MeV}\right)^{3/8}\\ \qquad\times \left(\dfrac{10^{-17}}{g}\right)^{5/8} %2.086412824e8 GeV
     \qquad \text{(quadratic).}
\end{array}\right.
\ee 
This could be violated if the dark Higgs took a statistically exceptional value in our Hubble patch during inflation. 

For the quartic and hybrid regimes,  a small coupling $\lambda$ satisfying Eq.~(\ref{eq:isoParam}) ensures that there is a value of $H_I$ consistent with both limits. However, a region of parameter space in the quadratic case is favored by
the combination of (\ref{eq:HIupperBound}) and (\ref{eq:HIlowerBound}),
\be \label{eq:likelihoodQuad}
m_{A'} \lesssim 56\ \text{keV}\ f_{A'}^{-6/5}\left(\frac{10^{-20}}{\lambda}\right)^{3/10}\!\!\!\left(\frac{g}{10^{-17}}\right)^{3/5} \text{(quadratic).}
\ee 
Comparing with Eq.~(\ref{eq:regimesmA}), this constraint rules out almost all of the quadratic regime for $f_{A'}=1$ and $\lambda=10^{-20}$.

\begin{figure}[!t]
\centerline{
\includegraphics[scale=0.85]{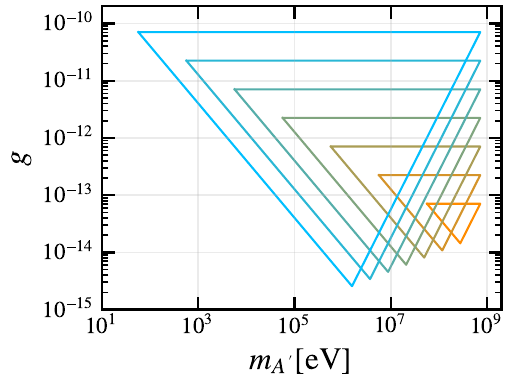}
}
\vspace{-0.2cm}
\caption{The allowed region for producing $A'$ as DM from dark $h$ decays in $\lbrace m_{A'},g\rbrace$ space is shown for different values of the dark Higgs self-coupling $\lambda$.  The value of $\lambda$ ranges from $10^{-20}$ (light blue) to $10^{-26}$ (orange) and decreases by a factor of 10 between consecutive contours.
\label{fig:mg3d}}
\end{figure}

We have now derived the main constraints that lead to a preferred
region of dark gauge coupling $g$ versus dark photon mass $m_{A'}$, shown 
in Fig.\ \ref{fig:mg} as the white triangle.  At a fixed value of $\lambda = 10^{-20}$, near the maximum allowed by isocurvature constraints, $g \lesssim 10^{-10}$ is required for $h\to A'A'$ to be kinematically allowed, Eq.\ (\ref{massratio}).  The lower left region is excluded by the coldness constraint  (\ref{eq:coldnessBound}) on $A'$, while the lower right region is limited by needing an initial Higgs displacement that is consistent with isocurvature constraints, Eq.\ (\ref{eq:likelihoodQuad}).  The region where the Higgs decays while oscillating in the quartic part of its potential (upper left) is excluded.  Smaller values of $\lambda$ require lower values of $g$ to satisfy the criterion $m_{A'}<m_h/2$ and they would delay the dark Higgs decay, making dark photons non-relativistic at a later time and strengthening the coldness constraint. As $\lambda$ decreases, the allowed triangle shrinks as shown in Fig.\ \ref{fig:mg3d}, disappearing when $\lambda \lesssim 10^{-27}$.

\section{Other constraints} \label{sec:inf}
In previous sections, we have derived a number of requirements for inflationary dark Higgs misalignment to be able to produce the desired relic density of dark photons through $h\to A'A'$ decay.  Here we describe further self-consistency checks, and demonstrate that other production mechanisms are subdominant in the favored region of parameter space that we will identify.

\subsection{Inflationary scale}

The stochastic misalignment mechanism requires the dark Higgs to be lighter than the Hubble scale during inflation; otherwise it evolves towards the minimum of the potential prematurely, diluting the dark sector energy density. However, the isocurvature bound discussed in section \ref{sec:misalignment} independently requires $h$ to be light during inflation. This is explicit in Eq.~(\ref{eq:isoParam}) for the quadratic case. In the quartic regime,  $m_h\sim\sqrt{3\lambda}h_0$ during inflation. Combining the condition $m_h<  H_I$ with the misalignment bound $h_0\gtrsim5.2\times10^4 H_I$ (see Eq.~(\ref{eq:isoField})) gives the constraint $\lambda\lesssim10^{-10}$, in agreement with Ref.~\cite{Dror:2018pdh}. This is much milder than the upper bound $\lambda\lesssim\mathcal{O}(10^{-20})$ obtained in Eq.~(\ref{eq:isoParam}), which takes the statistical distribution of $h_0$ across different Hubble patches into account. Therefore, the dark Higgs is consistently frozen everywhere in our allowed $\lbrace m_{A'},g\rbrace$ parameter space.

Eqs. (\ref{eq:phi0DM}) and (\ref{eq:phi0DMquad}) indicate that the dark Higgs amplitude must be large in order to produce the desired relic density, especially in the quartic scenario. The energy density of the scalar field must be subdominant during inflation, $\rho_\phi< \Lambda_I^4$, where $\Lambda_I$ is the inflationary energy scale, otherwise after reheating the Universe would be dominated by the dark sector rather than SM radiation. Using $H_I^2 = \Lambda_I^4/(3M_p^2)$ and assuming that $A'$ constitutes all the DM,  Eq.~(\ref{eq:HIupperBound}) implies that this energy condition is satisfied. For $\lambda= 10^{-20}$, we obtain the weak constraint $m_{A'}\gtrsim5\times10^{-13}$ eV in the quartic regime and $m_{A'}\lesssim 5g\,M_p$ in the quadratic scenario, both of which are satisfied everywhere in the region of interest.

Moreover, the {\it Planck} collaboration constrains $\Lambda_I\lesssim1.6\times10^{16}$ GeV for the inflationary scale \cite{Planck:2018jri}, hence $H_I\lesssim6\times 10^{13}$ GeV. Eq.~(\ref{eq:HIlowerBound}) then bounds the  dark photon's mass: $m_{A'}\gtrsim10^{-4}$ eV for the quartic regime and $m_{A'}\lesssim(g/10^{-17})^{5/3}\times10^{11}$ GeV,  using $\lambda=10^{-20}$ still. Again, both limits are less stringent than the ones we previously derived. We therefore see that the additional constraints imposed by inflationary dynamics are satisfied.

\subsection{Weak gravity conjectures}
The hierarchy $m_{A'}\ll m_h$ and the isocurvature constraint $\lambda \ll1$ lead us to favor small values of the gauge coupling $g\lesssim 10^{-10}$. In the limit $g\to0$, the $U(1)$ gauge symmetry becomes global, which is widely believed to be inconsistent with gravitational effects \cite{Kallosh:1995hi}. 

The weak gravity conjecture (WGC) \cite{Arkani-Hamed:2006emk} imposes a lower bound on the gauge coupling of an Abelian symmetry. In its simplest version, it requires the existence of at least one charged species with mass $m\lesssim gM_p$ to allow for the evaporation of near-extremal black holes. For most of the parameter space, the dark Higgs itself satisfies $m_h\lesssim g M_p$, namely for dark photon masses
\be 
m_{A'}\lesssim 17\ \text{MeV}\, \left(\frac{g}{10^{-15}}\right)^2 \left(\frac{10^{-20}}{\lambda}\right)^{1/2}. %17.21805012 MeV
\ee
Dark photons that violate this bound are already ruled out by the nonrelativistic constraint Eq.~(\ref{eq:coldnessBound}) or the likelihood limit Eq.~(\ref{eq:likelihoodQuad}), unless they only represent a small fraction of the total dark matter.

In our scenario, the $U(1)$ gauge symmetry is in its Higgs phase from the start. In this phase, it is often argued that, since charge is not conserved, interactions between an extremal black hole and the background field allows it to radiate \cite{Redi:2022zkt, Harlow:2022ich}. The WGC should only be enforced in the UV limit where the gauge symmetry is restored. We will argue in the next subsection that, because of the feeble couplings we are considering, the dark sector does not thermalize, let alone reach temperatures $T>v$. One could  imagine that the $U(1)$ is part of a larger symmetry group in the UV which could satisfy the WGC. 

To bolster this argument, notice that a light Dirac fermion $\psi$ charged under $U(1)$ could  be added to our framework without changing the phenomenology. As long as it is heavier than $m_{A'}/2$, dark photons would remain stable. Gauge symmetry prevents a direct Yukawa interaction $h \bar{\psi}\psi$, and the dark Higgs can produce fermions in the early universe only by the three-body decay $h\to A'A^{\prime*}\to A' \bar{\psi}\psi$, whose rate is suppressed by an extra factor of $g^2$ relative to $h\to A'A'$. Hence such dark fermions would not be be significantly produced, but they could allow extremal black holes to evaporate in regions where the dark Higgs cannot. In light of these arguments, and the relative weakness of the WGC bound on Fig.\ \ref{fig:mg} (shown as a green dashed line), the WGC does not seem worrisome. 

A related constraint is known as the sublattice weak gravity conjecture (sWGC) \cite{Heidenreich:2016aqi}, which imposes a UV cutoff at a scale $\Lambda\lesssim g^{1/3} M_{p}$. Our framework is consistent if it is higher than the Hubble parameter, $H_I\lesssim g^{1/3} M_{p}$. As explained in section \ref{sec:dm}, a smaller $H_I$ makes it less likely for the initial scalar field value to yield the dark matter relic density. Combining the sWGC with our lower bound on $H_I$ Eq.~(\ref{eq:HIlowerBound}) yields 
\be 
m_{A'} \left\lbrace \begin{array}{l}
     \gtrsim 3.6\times 10^{-4}\ {\rm eV}\ f_{A'}\left(\dfrac{\lambda}{10^{-20}}\right)^{5/8}\left(\dfrac{10^{-15}}{g}\right)^{1/2}\\ \quad \text{(quartic, hybrid)}, \\%3.563872997e-4 eV
     \lesssim 5.4\times 10^8\ {\rm GeV}\ f_{A'}^{-2/3}\left(\dfrac{10^{-20}}{\lambda}\right)^{5/6} \left(\dfrac{g}{10^{-17}}\right)^{23/9}\\ \quad \text{(quadratic)}.%5.426071113e8 GeV
\end{array}\right.
\ee 
These limits, shown as purple on Fig.\ \ref{fig:mg}, are less stringent than the coldness and likelihood constraints.

\begin{figure*}[!t] 
\centerline{
\includegraphics[scale=0.8]{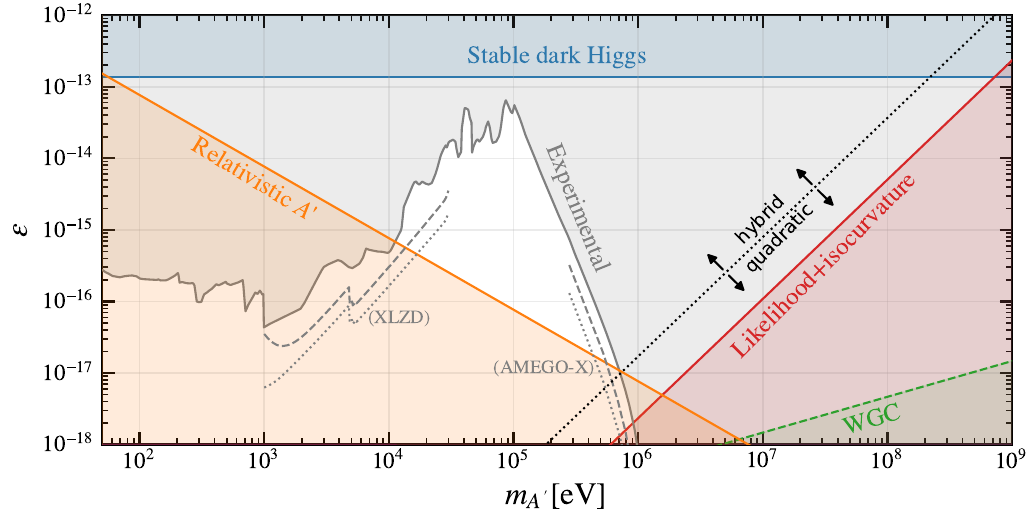}
}
\vspace{-0.2cm}
\caption{Constraints on kinetic mixing versus dark photon mass,
 assuming the one-loop estimate Eq.~(\ref{eq:epsilon}) for $\epsilon$. Grey: experimental constraints adapted from Refs.~\cite{Caputo:2021eaa,Caputo:2026pdw} (solid) and projected sensitivities of XLZD and AMEGO-X assuming realistic (dashed) and optimistic (dotted) detector efficiencies. %
 The other constraints are the same as in Fig.~\ref{fig:mg}. \label{fig:meps}}
\end{figure*}

\subsection{Competing production mechanisms} \label{sec:gravProd}
In this part, we will discuss other possible ways of producing dark photons, and show that they are not relevant at the small couplings needed for the mechanism of interest in this work.
\medskip

{\bf Gravitational Production.}
Unlike scalar particles, vectors cannot be produced directly by the misalignment mechanism unless a nonminimal coupling to gravity is introduced.  Ref.~\cite{Graham:2015rva} showed that quantum fluctuations of longitudinal modes of a vector mediator grow during inflation, resulting in a simple and efficient origin for dark photons. Assuming a Stueckelberg  mass term, this gravitational production mechanism yields a dark photon relic density
\be \label{eq:gravProdStuck}
f_{A',{\rm grav}}=\left(\frac{\Omega_{A'}}{\Omega_{\rm DM}}\right)_{\rm grav} \simeq \left(\frac{m_{A'}}{6 \mu\text{eV}}\right)^{1/2}\left(\frac{H_I}{10^{14}\ {\rm GeV}}\right)^2.
\ee 

Given the upper bound set by Planck on $H_I$,
Eq.\ (\ref{eq:gravProdStuck}) could work for $A'$ as light as $m_{A'}\sim10^{-5}$ eV. However, Stueckelberg dark photons are constrained by a magnetic version of the WGC, which in its weakest form requires a UV cutoff $\Lambda\lesssim m_{A'}^{1/5}M_p^{4/5}$ \cite{Reece:2018zvv}. Self-consistency requires this cutoff be above $H_I$ which in combination with Eq.~(\ref{eq:gravProdStuck}), implies that $m_{A'}\gtrsim0.3$~eV.

Gravitational production of Higgsed dark photons is less efficient because the $A'$ mass becomes time-dependent. Ref.~\cite{Sato:2022jya} studied a general model assuming power-law dependence 
 $m_{A'}\propto a^{2p}$ after inflation, until some critical point where $h$ settles to its VEV and $m_{A'}$ becomes constant. In the quartic and hybrid regimes, $p=1/2$ since $h\propto a^{-1}$ before the decays,  whereas in the quadratic regime $m_{A'}$ is constant. Then
 \begin{equation}
f_{A',{\rm grav}}\simeq
\begin{dcases}
   \left(\dfrac{m_{A'}}{gh_0}\right)^2 \left(\dfrac{m_{A'}}{6 \mu\text{eV}}\right)^{1/2}\left(\dfrac{H_I}{10^{14}\ {\rm GeV}}\right)^2 
   &\\ \quad\text{if }gh_0<\tilde m; \\[1ex]
   \left(\dfrac{m_{A'}}{0.8\ \text{GeV}}\right) \left(\dfrac{H_I}{10^{14}\ {\rm GeV}}\right)^{3/2}
   &\\ \quad\text{if }gh_0>\tilde m,
\end{dcases}
\label{eqfA}
\end{equation}
where $gh_0$ is the effective dark photon's mass during inflation  and $\tilde m=(m_{A'}H_I^{3})^{1/4}$. 

Using Eqs.~(\ref{eq:HIupperBound}) and (\ref{eq:HIlowerBound}) to estimate $H_I$ in Eq.\ (\ref{eqfA}) shows that the first case corresponds to the quadratic regime and a small part of the hybrid regime. For the quartic and most of the hybrid regime one should use the lower case of
Eq.\ (\ref{eqfA}). For quadratic oscillations, $gh_0=m_{A'}$ and we recover the Stueckelberg relic density, Eq.~(\ref{eq:gravProdStuck}).
Using the upper bound (\ref{eq:HIupperBound}) on $H_I$ imposed by isocurvature perturbations,  gravitationally produced $A'$ make up a fraction $f_{A',{\rm grav}}\sim 10^{-11}\,\lambda^{1/4}/g^{1/2}$ in the quadratic regime and $\sim 10^{-8}\, \lambda^{1/4}$ in the quartic scenario. These are negligible for the parameters of interest, leaving perturbative decays of the misaligned Higgs as the dominant production mechanism.

{\bf Parametric resonance.}
In the quartic and hybrid regimes, $h$ periodically crosses $h=0$, at which point $A'$ becomes massless and violates the adiabaticity condition $\dot m_{A'} \ll m_{A'}^2$. This gives nonperturbative production of longitudinal $A'$s \cite{Dror:2018pdh} (see also \cite{Khan:2026php}). Focusing on a conformally invariant  $V(h)$, Ref.\  \cite{Dror:2018pdh} found that in the limit $g\ll \sqrt{\lambda}$, parametric resonance yields only $\Omega_{A'}/\Omega_{\rm DM}\sim g/\sqrt{\lambda}\ll1$, leaving undecayed $h$ as the dominant component of the dark sector. This applies directly to our hybrid regime, while parametric resonance is absent in the quadratic regime because $m_{A'}$ is nearly constant and the adiabaticity condition is always satisfied.

{\bf Cosmic string decays} can  produce dark photons \cite{Long:2019lwl,East:2022rsi}. To satisfy isocurvature constraints, the $h$ field, which has acquired a VEV $\gtrsim v$,  must be homogeneous over large distances, which effectively breaks the $U(1)$ symmetry before or during inflation and prevents string formation. Cosmic strings could form if the dark sector thermalizes to a temperature $T_D\gtrsim v$ after reheating, which restores the $U(1)$ symmetry. 

However, that is not a possibility in our setup, given the weakness of the couplings. We estimate 
$T_D$ by setting
the $A'A'\to A'A'$ elastic scattering rate $\Gamma_{\rm el}\approx \alpha_D^2(m_{A'}/m_h)^2 T_D$  equal to the Hubble rate, achieving kinetic equilibrium. Here, $\alpha_D=g^2/4\pi$ and the mass suppression factor comes from an internal $h$ exchange. This gives
\be 
T_D \sim \alpha^2_D r^2 \left(\frac{m_{A'}}{m_h}\right)^2 M_p = \frac{g^6 r^2}{16\pi^2 \lambda} M_p, 
\ee 
where $r\equiv T_D/T$ is the ratio of the dark sector and SM temperatures. Our preferred values of $g$ and $\lambda$ imply $T_D<10^{-25}$ GeV,
hence dark photons never reach kinetic equilibrium. Chemical equilibrium would require number-changing processes such as $2\to3$ and $2\to4$ scatterings, which are further suppressed by additional gauge insertions. 

We also do not expect $h$ to reach kinetic equilibrium prior to its decay as it behaves like a coherently oscillating field, not a collection of scattering particles. Even if a non-negligible population of $h$ particles were produced (e.g. via parametric resonance or freeze-in), we can similarly estimate its elastic scattering rate as $\Gamma_{\rm el}\approx (\lambda/4\pi)^2 T_D$, resulting in a temperature $T_D<10^{-24}$~GeV at equilibrium. Hence the dark sector never thermalizes and the $U(1)$ symmetry remains broken after inflation.

{\bf Freeze-in.}  Another possible production mechanism is freeze-in from SM scattering. The standard scenario involves kinetic mixing between visible and dark photons, which will be discussed in section \ref{sec:epsilon}. In the absence of kinetic mixing, freeze-in can  occur by graviton exchange~\cite{Garny:2015sjg}, which is Planck-suppressed and only efficient for high reheating temperatures $T_R$. Ref.~\cite{Redi:2020ffc} estimated the resulting abundance for a conformal dark sector to be
\be 
Y_D\sim 10^{-5} \left(\frac{T_R}{M_p} \right)^3 \sim \frac{5\times 10^{-5}}{g_*^{3/4}}  \left(\frac{H_I}{M_p} \right)^{3/2},
\ee 
assuming instantaneous reheating, $H_I\approx 0.3 g_*^{1/2} T_R^2/M_p$. Using  Eq.~(\ref{eq:HIupperBound}), we find $Y_D\sim 10^{-17}\times (1\ {\rm eV}/m_{A'})$, much lower than the abundance $Y_D\sim 0.4\times (1\ {\rm eV}/m_{A'})$ needed for the dark matter relic density \cite{Cline:2018fuq}. Thus, graviton exchange is not relevant in our scenario.

\section{Constraints on kinetic mixing}
\label{sec:epsilon}
Thus far we have ignored possible portals between the hidden sector and the standard model.  If a heavy particle
$\chi$
carrying both SM hypercharge and the dark U(1) exists, it induces 
a low-energy kinetic mixing Lagrangian
\be 
\mathcal{L}_{\rm mix} = \frac{\epsilon}{2} F_{\mu \nu }F^{\prime \mu \nu}.
\ee
at one loop, 
with mixing parameter  of order
\be \label{eq:epsilon}
\epsilon = q_g q_e\, \frac{ ge}{16\pi^2} \log\left(\frac{m_\chi^2}{\Lambda^2}\right)\,,
\ee 
assuming $\epsilon$ was zero at some high scale $\Lambda > m_\chi$. Here $q_g$ and $q_e$ are the dark $U(1)$ and SM electric charges of $\chi$, which we set to 1 for simplicity.

In this section, we will assume that
inflation occurs at $H_I < m_\chi$ and therefore $\epsilon$ is not running.
For illustration, we set the log to unity in Eq.\ \ref{eq:epsilon} so that
$\epsilon = g e/(16\pi^2)$.  The constraints we derived in Fig.\ \ref{fig:mg} can then be simply transcribed to the $m_{A'}$-$\epsilon$ plane,
which are shown in Fig.~\ref{fig:meps}.
There are now additional experimental constraints specific to $\epsilon$ that must be taken into account, and which significantly reduce the previously allowed parameter space.

Dark matter stability requires its lifetime to be greater than the age of the Universe, $\tau=\Gamma^{-1}\gtrsim4.3\times 10^{17}$ s, or $\Gamma\lesssim1.5\times10^{-42}$\,GeV. If $m_{A'}>2 m_e$, dark photons can decay into SM electron-positron pairs through kinetic mixing, with rate
$\Gamma_{e\bar{e}}  \simeq  (\alpha/3) \epsilon^2 m_{A'},$
 assuming $m_{A'}$ is not too close to the kinematic threshold.  At smaller $m_{A'}$, there 
 are decays into three photons with rate $
\Gamma_{3\gamma} \simeq 4.7\times 10^{-8}\ \alpha^4 \epsilon^2 ({m_{A'}^9}/{m_e^8})$
\cite{Caputo:2026pdw}.  
Indirect detection of this radiation typically give stronger constraints on $\epsilon$ than the
lifetime limit \cite{Linden:2024fby,Jeesun:2026ryo},  which are taken into account in the grey ``Experimental'' excluded region on
Fig.\ \ref{fig:meps}.
 
Dark Higgs emission in stars can occur via plasmon decays, which would be an anomalous cooling mechanism. The decays are kinematically blocked if $m_h\gtrsim1$ keV, the core temperature of stars. This motivated Ref.~\cite{Cline:2024wja} to focus on large coupling $\lambda\approx8\pi/3$ in order to maximize the dark Higgs mass. However, this overlooked the isocurvature constraints, that require $\lambda \lesssim 10^{-20}$ in a scenario consistent with the misalignment mechanism. Consequently, there is a lower bound on the VEV $v$, or alternatively on the free parameters $m_{A'}$ and $\epsilon$,
\be 
m_{A'} \gtrsim 3.7\,% 3.6873735 eV
{\rm eV}\ \left(\frac{10^{-20}}{\lambda}\right)^{1/2} \left(\frac{\epsilon}{10^{-15}} \right).
\label{HBbound}
\ee
This is a sufficient but not necessary condition for evading stellar cooling by Higgsstrahlung \cite{An:2013yua}.  If the dark Higgs boson is lighter than $\sim 1$\,keV, the stellar cooling limit is $g\epsilon < 10^{-14}$.  Using the naturalness estimate Eq.\ (\ref{eq:epsilon}) for $\epsilon$ , this translates to 
$\epsilon < 4.4\times 10^{-9}$,
which is satisfied in the region of Fig.\ \ref{fig:meps}.  Hence the stellar cooling bounds are not relevant for us.

For consistency with the assumption that the dark photon abundance is
generated predominantly by $h$ decays, production induced by kinetic
mixing with the SM plasma must remain subdominant. This includes
nonresonant freeze-in through processes such as
$e\gamma\rightarrow eA'$, as well as resonantly enhanced conversion of
in-medium photons into dark photons. Ref.~\cite{Cline:2024wja} showed
that the nonresonant process $e\gamma\rightarrow eA'$ is inefficient for
\[
\epsilon\lesssim
10^{-8}\left(\frac{\mathrm{eV}}{m_{A'}}\right)^{1/2}.
\]
A stronger constraint in part of the parameter space follows from
resonant $\gamma\rightarrow A'$ conversion
\cite{Mirizzi:2009iz,Redondo:2008ec,Arias:2012az}.

In the nonadiabatic limit the conversion probability is
\[
P_{\gamma\to A'}=
\frac{\pi\epsilon^2m_{A'}^4}
{\omega\,|dm_\gamma^2/dt|},
\]
which gives
\[
n_{A'}=
\frac{\pi\epsilon^2m_{A'}^2T_{\rm res}^2}
{6pH_{\rm res}}
\]
when $m_\gamma^2\propto T^p$. For a relativistic $e^\pm$ plasma, $p=2$ and
$T_{\rm res}\simeq 8$--$10\,m_{A'}$; requiring that this population
not exceed the observed dark-matter density gives
$\epsilon\lesssim2\times10^{-12}$ for
$m_{A'}\gtrsim50~{\rm keV}$.

At lower masses the resonance occurs during $e^\pm$ annihilation,
where $m_\gamma(T)$ is not accurately described by a single power
law.  The numerical calculation of
Ref.~\cite{Redondo:2008ec} gives a mixing of order $\sim 10^{-12}$ for
resonant production to account for all of the dark matter throughout
the $10$--$50~{\rm keV}$ interval relevant to Fig.~\ref{fig:meps}.
Since the viable region has $\epsilon\lesssim{\rm few}\times10^{-14}$,
the resonantly produced component is negligible and production from $h$ decays dominates.

For completeness, well after $e^\pm$ annihilation one has
$T_{\rm res}\simeq3~{\rm keV}
(m_{A'}/{\rm meV})^{2/3}$, valid for
$m_{A'}\lesssim15~{\rm meV}$.  In this regime even complete conversion of the photon bath gives only $\Omega_{A'} /\Omega_{\rm DM}\lesssim
m_{A'}/(3~{\rm eV})\ll1$, so no dark-matter overproduction bound arises.

\subsection{Projected sensitivity to dark photon absorption at XLZD}

Our allowed region is constrained from above in
Fig.\ \ref{fig:meps} by experimental searches, which give prospects for its discoverability, or falsifiability.
Here we estimate the reach of next-generation liquid xenon observatories such as XLZD \cite{XLZD:2024nsu} to dark photon dark matter in the mass window $m_{A'}\!\sim\!1$-$30\,$keV, which is relevant for our predictions.

A nonrelativistic $A'$ of the local dark matter population is absorbed on bound
electrons in complete analogy with the photoelectric effect for an ordinary
photon of energy $\omega=m_{A'}$ \cite{Pospelov2008, An2015, Bloch2017}. The event rate is
\begin{equation}
  R(m_{A'}) \;=\; \mathcal{E} \, \eta (E) \, \varepsilon^{2}\,
  \frac{\rho_{\rm DM}}{m_{A'}}\, 
  \frac{\sigma_{\rm pe}(m_{A'})}{m_{\rm Xe}} \, ,
  \label{eq:rate}
\end{equation}
where $\rho_{\rm DM}=0.3\,{\rm GeV\,cm^{-3}}$, the dark matter local energy density \cite{Read:2014qva}, $m_{\rm Xe}$ is the xenon atomic mass, and $\sigma_{\rm pe}$ is the atomic photoabsorption cross section \cite{Henke1993}. $\mathcal{E}$ denotes the experiment exposure.

For an exposure $\mathcal{E}=M\,T$, with  detector mass $M$ and  live-time $T$ of the experiment, and detection efficiency $\eta (E)$ (taken from XENONnT \cite{XENON:2022ltv}), the
monoenergetic absorption line is smeared by the detector energy resolution
$\sigma_E/E \simeq 0.31/\sqrt{E/{\rm keV}}$ into $0.1\,$keV bins \cite{XENON:2020iwh}. We derive the $90\%$~C.L.\ median expected upper limit from the binned Asimov
profile likelihood~\cite{Cowan2011}, solving
\begin{equation}
  q_A(\varepsilon^2)=2\sum_i\!\left[\varepsilon^2 s_i
   - b_i\ln\!\Big(1+\tfrac{\varepsilon^2 s_i}{b_i}\Big)\right]
   =\big[\Phi^{-1}(0.9)\big]^2 .
  \label{eq:asimov}
\end{equation}
 Here $\varepsilon^2 s_i$ is the expected signal in bin $i$, with $s_i$ the
resolution-binned rate of Eq.~\eqref{eq:rate} evaluated at $\varepsilon^2=1$, and $b_i$ the expected background counts, which we take from XENONnT \cite{XENON:2022ltv}. Equation
\eqref{eq:asimov} is the background-only ``Asimov'' profile-likelihood ratio
$q_A=-2\ln[L(\varepsilon^2)/L(0)]$~\cite{Cowan2011}, obtained by setting the bin counts equal to their background expectation; this yields the median expected sensitivity. We solve Eq.~\eqref{eq:asimov} numerically at each mass $m_{A'}$. In the background-dominated regime ($\varepsilon^2 s_i\ll b_i$) it reduces to the bin-summed
$S/\sqrt{B}$ form, $\varepsilon^2=1.28\,(\sum_i s_i^2/b_i)^{-1/2}$,
which makes the scaling $\varepsilon\propto\mathcal{E}^{-1/4}$
explicit, leading to a modest improvement over current limits in Fig. \ref{fig:meps}

We bracket the sensitivity with two scenarios in Fig.~\ref{fig:meps}: a conservative case ($\mathcal{E}=200\,$t$\cdot$yr with the full background above and the real efficiency) and an optimistic case
($\mathcal{E}=1000\,$t$\cdot$yr with only the solar-$pp$-neutrino floor and perfect
efficiency). The projected $90\%$~C.L.\ reach on the kinetic mixing is roughly one order of magnitude below current bounds and probes the upper portion of the allowed region of our model.

\subsection{Projected sensitivity to $A^{\prime}$ tridents with AMEGO-X}
For $m_{A'}<2m_e$, the dominant visible decay channel for dark photons is $A'\to3\gamma$, with width \citep{Pospelov2008b,McDermott2018}
\begin{equation}
\Gamma_{3\gamma}\simeq
1\,{\rm s^{-1}}\,(\epsilon/0.003)^{2}\,(m_{A'}/m_e)^{9}.
\end{equation}
The 3-photon energy spectrum is \cite{Linden2024}
\begin{equation}
dN/dx=\tfrac{2}{17}x^{3}(1715-3105x+\tfrac{2919}{2}x^{2}),
\end{equation}
with $x=2E_\gamma/m_{A'}$. The Galactic signal intensity is
\begin{equation}
\frac{d\Phi}{dE_\gamma\,d\Omega}=\frac{\Gamma_{3\gamma}}{4\pi\,m_{A'}}\,
\frac{dN}{dE_\gamma}\,\mathcal{D},
\end{equation} 
with $\mathcal{D}$ the dark matter column density for an NFW profile with normalization anchored at the solar system as $\rho_\odot=0.3\,{\rm GeV\,cm^{-3}}$, and scale radius $r_s=20\,$kpc, then averaged over the
region $|\ell|,|b|<47.5^\circ$. We find $\mathcal{D}\simeq 4.3 \times 10^{22}$GeVcm$^{-2}$. We derive the AMEGO-X sensitivity projections by requiring that this intensity not exceed $2\,f_{\rm unc}\,B(E_\gamma)$, where $B$ is the diffuse astrophysical background (cosmic X-ray background
plus Galactic emission) and $f_{\rm unc}$ the fractional precision to
which it is determined. For INTEGRAL/SPI, $f_{\rm unc} \simeq 15\%$ \cite{Siegert2022}, and our prescription reproduces the limit from \cite{Linden2024} within a factor of $\sim 2$. For AMEGO-X~\cite{Fleischhack:2021mhc,AMEGOX2022}, whose larger effective
area and strongly suppressed instrumental background may allow a more
precise determination of the same diffuse sky, we assume
$f_{\rm unc}=5\%$ for the 3-year baseline mission (conservative) and
$f_{\rm unc}=1.5\%$ for a $\sim$10-year extended mission (optimistic,
statistics-limited). The resulting projections apply for
$m_{A'}\gtrsim0.3\,$MeV, where the spectral peak lies above the
AMEGO-X threshold of $100\,$keV, and are shown in Fig.~\ref{fig:meps}. This would probe currently allowed parameter space for $m_{A'}\simeq0.3$-$0.8\,$MeV, reaching the boundary of the region
consistent with our production mechanism at its high-mass end and
complementing the XLZD projection at low masses.

\section{Conclusions}
\label{sec:conc}

In this work we carefully studied a simple mechanism for producing dark photons as the dark matter of the Universe: the dark Higgs that gives mass to $A'$ is excited during inflation and subsequently decays as $h\to A'A'$.  Although the mechanism requires small $h$ self-couplings $\lambda\lesssim 10^{-20}$ to satisfy isocurvature constraints, such values are technically natural \cite{tHooft:1979rat} since the gauge coupling is also small, $g\lesssim 10^{-10}$.  We identified a significant region of parameter space consistent with this mechanism, for $100\,{\rm eV}\lesssim m_{A'}\lesssim 1\,$GeV,
and also consistent with the weak gravity conjecture, despite the smallness of the gauge coupling.

The scenario becomes somewhat more constrained when one assumes a generic level of kinetic mixing $\epsilon \sim g e/(16\pi^2)$, leading to a narrower range of allowed masses from 10\,keV to 1\,MeV, but an interesting parameter region remains viable.  This includes the potential for discovery in upcoming direct detection experiments, like XLZD.

This model makes several predictions that can be tested by future astronomical and cosmological observations. Eqs.~(\ref{eq:HIupperBound}) and (\ref{eq:HIlowerBound}) indicate that the preferred parameter space corresponds to $H_I\sim10^{5}-10^{10}$ GeV. Such values would imply a tensor-to-scalar ratio $r=A_t/A_s\sim(H_I/M_p)^2/A_s\lesssim10^{-8}$, well below the projected sensitivities $r\sim10^{-3}$ of upcoming CMB polarization experiments such as LiteBIRD \cite{LiteBIRD:2022cnt,LiteBIRD:2024wix}. Consequently, the detection of primordial tensor modes in the CMB would rule out the simplest realization of this mechanism.

Future measurements of isocurvature fluctuations provide a complementary test. The analysis presented in Appendix \ref{app:iso} implies that the upper bound on the dark Higgs self-coupling Eq.~(\ref{eq:isoParam}) scales as $\lambda\propto\beta_{\rm iso}^{1/2}$. Future CMB observations are expected to improve the sensitivity to primordial isocurvature perturbations by several-fold \cite{Bellomo:2022qbx},  probing quartic couplings within a factor of a few of the present bound.

The lower edge of the allowed dark photon mass range is determined primarily by structure formation, requiring the dark photons to cool sufficiently before the formation of galaxies. Future surveys of the Lyman-$\alpha$ forest such as DESI \cite{DESI:2024lzq}, WEAVE-QSO \cite{WEAVE:2016rxg} and Euclid \cite{Euclid:2019clj} are expected to improve the sensitivity to warm dark matter and could therefore strengthen the lower bound on $m_{A'}$. Finally, future gamma-ray observatories like AMEGO-X and COSI \cite{Tomsick:2023aue} should substantially improve measurements of the diffuse soft gamma-ray spectrum down to energies of $\sim0.1-0.2$ MeV. If dark photons possess a generic one-loop kinetic mixing, these observations could improve the present constraints from the $A'\to 3\gamma$ decays below the $e^+ e^-$ threshold.

\bigskip
{\bf Acknowledgments.}  The research of JC and JSR 
is supported by the Natural Sciences and Engineering Research Council (NSERC) of Canada.
The work of GH was supported by the Neutrino Theory Network Fellowship with contract number 726844, and by the U.S. Department of Energy under award number DE-SC0020262. This manuscript has been authored by FermiForward Discovery Group, LLC under Contract No.
89243024CSC000002 with the U.S. Department of Energy, Office of Science, Office of High Energy Physics.

\appendix

\section{Initial field value distribution function}
\label{app:PDF}
Here we derive formulas for the mean values and standard deviations of the stochastically generated
dark Higgs VEV at the end of inflation.
Stochastic misalignment generates values of the complex scalar field $H=h e^{i\theta}/\sqrt2$ inside a Hubble patch that are randomly distributed according to Eq.~(\ref{eq:pdfcomplex}),
\be 
f(H) \propto \mathrm{exp}\!\left(-\frac{8 \pi^2 \lambda}{3 H_I^4} \left(|H|^2-\frac{1}{2}v^2\right)^2 \right).
\ee 
We are interested in the distribution of the radial mode $h$. After integrating over the angular component $\theta$, the probability distribution (PDF) is
\be 
f(h) = \mathcal{N}\ h\, \exp\!\left(- \frac{(h^2-v^2)^2}{\Delta_h^4}\right), \quad  h>0,
\ee 
where $\mathcal N$ is a normalization factor and the characteristic width is $\Delta_h=[3H_I^4/(2\pi^2 \lambda)]^{1/4}$. We will consider the two limits where $\Delta_h\gg v$ and $\Delta_h\ll v$.

When $\Delta_h\ll v$, the exponential nearly vanishes everywhere except for a narrow region about $h=v$. Expanding $h=v+\phi$ and keeping only the leading contribution gives
\be 
f(\phi) \simeq \mathcal{N}\ {\exp}\!\left(- \frac{4v^2 \phi^2}{\Delta_\Phi^4}\right) = \mathcal{N}\ {\exp}\!\left(- \frac{4\pi^2 m_h^2}{3H_I^4}\phi^2\right), 
\ee 
where the near-constant prefactor $h\sim v$ has been included in the normalization constant. This is a Gaussian distribution centered around $\phi=0$ (\textit{i.e.,} $h=v$) with standard deviation
\be \label{eq:sigmaQuad}
\sigma = \sqrt{\frac{\Delta_h^4}{8v^2}} = \sqrt{\frac{3 H_I^4}{8\pi^2 m_h^2}}\qquad \text{(quadratic)}.
\ee 
Hence  $\abs{\phi}\lesssim 1\sigma$ with  68.3\%
probability, and  $\abs{\phi}\lesssim 2\sigma$   with 95.4\% probability.

In the opposite limit $\Delta_h\gg v$, we use the dominant quartic term in the exponential,
\be 
f(h) \simeq \mathcal{N}\ h\, \exp\!\left(- \frac{h^4}{\Delta_h^4}\right).
\ee 
This is a quartic distribution with prefactor $h$, which is less familiar. The normalization constant is obtained by requiring $\int_0^{\infty} f(h) dh=1$, giving
\be 
\mathcal{N}\simeq \frac{4}{\sqrt{\pi} \Delta_h^2}\,.
\ee 
Because of the prefactor $h$, 
the distribution is peaked at $h=\Delta_h/\sqrt{2}$ instead of zero. The value mean is
\be 
\ev{h}= \frac{\Gamma(3/4)}{\sqrt{\pi}} \Delta_h \simeq 0.69\, \Delta_h, %0.6913673390...
\ee
and its standard deviation is
\begin{align} 
\sigma &= \left(\ev{h^2}-\ev{h}^2\right)^{1/2} = \left(\frac{1}{\sqrt{\pi}} - \frac{\Gamma(3/4)^2}{\pi}\right)^{1/2}\Delta_h \nn\\
&\simeq 0.29\, \Delta_h \qquad {\rm (quartic)}. %0.2935997038...
\end{align}
One can check that the probability for $h$ to be  within $\pm 1\sigma $ ($\pm 2 \sigma$) of its mean value is 65.2 \% (96.7 \%), similarly to a Gaussian distribution. %0.9849670428 ; 1.278566747

\section{Isocurvature power spectrum} \label{app:iso}
As detailed in section \ref{sec:dm}, the initial amplitude $h_0$ of the dark Higgs determines the dark photon relic density. As a consequence, sub-Hubble fluctuations of the scalar field during inflation lead to inhomogeneities in the local dark photon abundance, which would leave an imprint on the CMB. One must ensure that these isocurvature perturbations are sufficiently suppressed to satisfy the bounds set by Planck.

During inflation, the light scalar field acquires fluctuations of order $\delta h\sim H_I/2\pi$. These fluctuations are not directly observable in the CMB; instead, they are transferred to dark photons via the decay of the scalar condensate. Therefore, the relevant quantity for the isocurvature power spectrum is the dark photons energy perturbations,
\be \label{eq:DPfluctuations}
\frac{\delta \rho_{A'}}{\rho_{A'}} = \frac{\partial \ln \rho_{A'}}{\partial \ln h} \frac{\delta h}{h}= q_{\rm eff} \frac{H_I}{2\pi h},
\ee 
where the local response parameter $q_{\rm eff}\equiv\sfrac{\partial\ln\rho_{A'}}{\partial \ln h}$ describes how the dark Higgs amplitude impacts the final dark photon abundance \cite{Khan:2026bmf}.  In our model, the dark photon energy density $\rho_{A'}=m_{A'}n_{A'}$ is related to the initial dark Higgs abundance via $n_{A'}=2Y_h s_{\rm osc}$. Using Eq.~(\ref{eq:habundance}), one finds $q_{\rm eff}=2$ for the quadratic regime and $q_{\rm eff}=3$ for the quartic case.

Assuming dark photons make up a fraction $f_{A'}=\Omega_{A'}/\Omega_{\rm DM}$ of the dark matter content of the Universe, Eq.~(\ref{eq:DPfluctuations}) leads to an isocurvature power spectrum
\be \label{eq:isoPS}
\mathcal{P}_S=f_{A'}^2 q_{\rm eff}^2  \left(\frac{H_I}{2\pi h_0}\right)^2.
\ee 
The amplitude of this spectrum is constrained by Planck to $\beta_{\rm iso}=\mathcal{P}_S/(\mathcal{P}_S+\mathcal{P}_R)<0.038$, where $\mathcal{P}_R$ is the adiabatic power spectrum, whose measured amplitude is $2.10\times 10^{-9}$. Using Eq.~(\ref{eq:isoPS})  leads to the constraint $h_0\gtrsim1.7\times10^{4}\ f_{A'}q_{\rm eff}\,H_I$, which yields Eq.~(\ref{eq:isoField}) when applied to the quadratic and quartic regimes (in the former case, $h_0$ should be replaced by $\phi_0$).

\bibliographystyle{utphys}
\bibliography{refs.bib}
\end{document}